# Aberration measurement and correction on a large field of view in fluorescence microscopy

T. FURIERI,[1,2] D. ANCORA,[3] G. CALISESI,[3] S. MORARA,[4] A. BASSI,[3] S. BONORA[1,*]

[1]*National Council of Research of Italy, Institute of Photonics and Nanotechnology, via Trasea 7, 35131, Padova, Italy*
[2]*University of Padova, Department of Information Engineering, Via Gradenigo 6, 35131, Padova, Italy*
[3]*Politecnico di Milano, Department of Physics, piazza Leonardo da Vinci 32, 20133 Milan, Italy*
[4]*National Council of Research of Italy, Institute of Neuroscience, via Vanvitelli 32, 20129, Milan, Italy*

*Corresponding author: stefano.bonora@pd.ifn.cnr.it

**The aberrations induced by the sample and/or by the sample holder limit the resolution of optical microscopes. Wavefront correction can be achieved using a deformable mirror with wavefront sensorless optimization algorithms but, despite the complexity of these systems, the level of correction is often limited to a small area in the field of view of the microscope.**

**In this work, we present a plug and play module for aberration measurement and correction. The wavefront correction is performed through direct wavefront reconstruction using the spinning-pupil aberration measurement and controlling a deformable lens in closed loop. The lens corrects the aberrations in the center of the field of view, leaving residual aberrations at the margins, that are removed by anisoplanatic deconvolution. We present experimental results obtained in fluorescence microscopy, with a wide field and a light sheet fluorescence microscope. These results indicate that detection and correction over the full field of view can be achieved with a compact transmissive module placed in the detection path of the fluorescence microscope.**

## 1. INTRODUCTION

Adaptive Optics is a technology used to improve image quality when the optical system is subject to phase aberrations. In optical microscopy the aberrations can be induced by imperfections and misalignments of the illumination or detection optics. This becomes particularly relevant when considering the continuous development of new microscopy configurations and measurement protocols, such as clearing methods [1], [2] or customized sample holders [3], [4]. On the other hand, the inhomogeneities in the refractive index of the sample itself, can induce aberrations and reduce the contrast of the formed image. In order to perform high resolution and super-resolution imaging, a good knowledge of these aberrations and their correction are required.

Recently, many techniques have been developed to compensate for optical aberrations in microscopy [5], [6]. They either rely on the use of a deformable mirror or on the use of a liquid crystal spatial light modulator placed in the pupil, or in its image plane. An approach similar to astronomical adaptive optics with a closed loop control is not convenient in microscopy, because the samples might not have bright point sources for the wavefront sensor. For this reason, wavefront sensorless control has become a standard for AO in microscopy. Remarkable achievements have been obtained with two techniques: image sharpening [6], [7] and pupil segmentation [8], [9]. Image sharpening uses a merit function that displays a maximum when the aberrations are minimized. With this assumption, it is easy to apply some bias aberrations to estimate the wavefront error using an optimization algorithm and then to compensate for them. Pupil segmentation is based on a different approach: by means of a liquid crystal panel, the pupil is segmented, and the phase of each segment is tuned, maximizing the constructive interference, thus resulting in an increased signal intensity.

Although adaptive optics has demonstrated its benefits, several limitations are still hindering its widespread diffusion. The level of the correction depends on the sample structure and it is not measurable, together with the correction being limited to the isoplanatic patch, which is often smaller than the field of view. In addition, microscopes must be entirely redesigned to include adaptive optics, adding complexity and limiting its adoption in commercial systems.

Some of these problems have been recently approached, but a solution that deals with them together is not yet available. For example [10], [11], [12], propose methods that strongly simplify the system using an adaptive lens, instead of a deformable mirror, combined with optimization algorithms. This has brought to promising results in microscopy and ophthalmic imaging [7], [13].

The problem of limited isoplanatic correction has been addressed with different techniques. Laslandes et al. [14], and Popovic et al. [15] apply multi-conjugate adaptive optics, using two deformable mirrors. Other approaches, such as anisoplanatic correction with liquid crystals, promise to be effective [16]. Furthermore, the dependence on the sample structure and contrast has been analyzed and solved adding high spatial frequencies in the illumination path [17].

In this paper we present a method to achieve anisoplanatic aberration correction in the entire field of view of a widefield microscope. The method consists in a compact optical module, able to detect and compensate for the aberrations. The detection is based on spinning-pupil aberration measurement (SPAM) that reconstructs the wavefront phase in the different locations of the field of view [18]. This device controls an adaptive lens based on multiple piezoelectric actuators [10], which is installed into the same chassis of the SPAM module, in a compact device. The adaptive lens corrects the aberrations in one point in the field of view. The remaining residual aberrations, still present in the other regions of the field of view, are removed by deconvolution. This is possible because the corresponding point spread function is known, as the wavefront phase has been detected by SPAM over the entire sample.

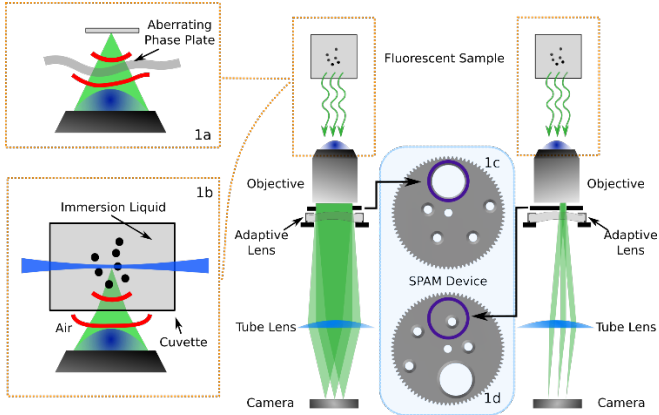

*Fig. 1. Optical layout of the imaging system used in this experiment including the SPAM module in combination with an adaptive lens. The diagram shows the SPAM module in the full pupil position (c) and in the sub-pupil position (d). The microscope has been used in wide field (a) and light sheet microscopy (b) configurations, as illustrated in the insets.*

## 2. RESULTS AND DISCUSSION

### 2.1 Integrated wavefront aberration measurement and adaptive optics correction

The method consists in a device, that incorporates both aberration detection and correction, placed in the proximity of a widefield microscope pupil (Fig1). The detection is performed using the spinning pupil aberration measurement (SPAM) device that, by moving an aperture inside the pupil and measuring the relative shift of each formed image, reconstructs the wavefront phase [18]. The correction is obtained using an adaptive lens (AL) made of ultrathin glass membranes bended by multiple piezoelectric actuators [10]. These lenses are compact and can be placed directly in the detection path of the optical microscope, which is in this case adjacent to the spinning sub-aperture module. The AL is controlled by the SPAM with a closed loop, as in typical adaptive optics applications [19] (see Material and Methods). The control system computes the actuators control voltages that minimize the wavefront deviation from flat using an integrative controller.

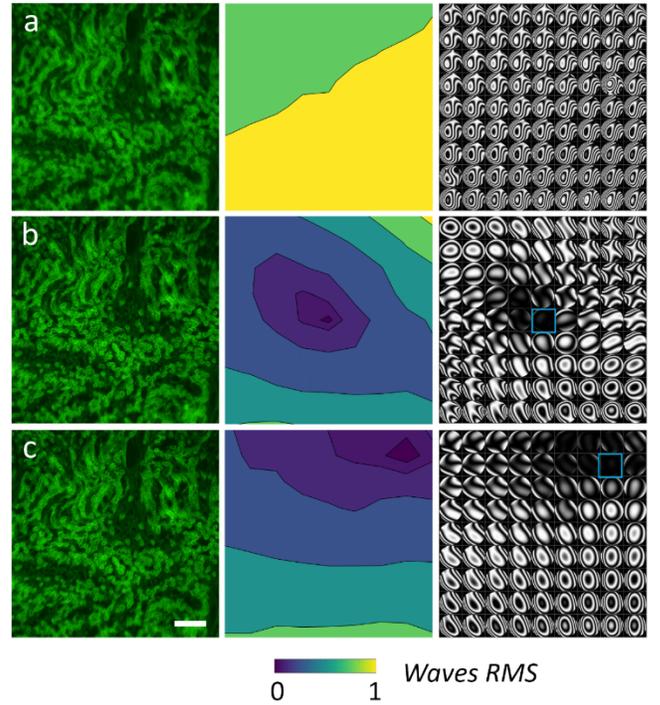

*Fig. 2. Comparison between the uncorrected and the AL corrected image, with correction in different region. Each raw shows an image of the sample, the RMS error (wavelength = 530 nm) and the wavefront phase aberrations represented as interferograms in a 9 x 9 grid. (a) Aberrated image without AL. (b) Corrected with AL in the central field. (c) Corrected with AL in the right-top corner. The correction region is indicated by the blue square. Scale bar is 200 µm.*

We tested this method in a wide field fluorescence microscope. The schematic of the system is shown in Figure 1 (fluorescence filters are not shown). The module (18 actuators Adaptive Lens with 10mm pupil, AOL1810 Dynamic Optics srl and spinning pupil) is placed in proximity of the back aperture of the objective lens (Mitutoyo, NA=0.28, 10X, long working distance). The effective Numerical Aperture of the system is 0.24, limited by the lens diameter. We imaged fluorescence sample slides (Fluocell sample prepared slice #3), with a mouse kidney section labeled with Alexa Fluor™ 488 WGA, with a phase plate, made of 1mm thick flexible transparent polystyrene plate, placed in front of it to induce further controlled amount of aberrations. In the first experiment, the lens starts in flat position. We measured the wavefront, aberrated by the phase plate, in 9x9 positions (Fig 2).

**Table 1: Comparison of the correction parameter for the wide field fluorescence microscope. The table reports the area (%) of the image below the Marechal criterion, the residual wavefront error for different corrected image segment (center and top-right, highlighted in blue in Fig. 2) and the average wavefront error on the whole image.**

|  | Below $\lambda/14$ (%) | Segment RMS error ($\lambda$) | Mean image RMS error ($\lambda$) |
|---|---|---|---|
| uncorrected | 0 |  | 0.78 |
| center | 3.7 | 0.04 | 0.33 |
| top-right | 8.6 | 0.03 | 0.30 |

The wavefront aberrations vary field to field with an average value of 0.78 waves, root mean squared (RMS), at 530 nm. In second place, we corrected the wavefront in a single point, at the center of the of the image. We then repeated the measurement correcting in the top-right corner. In both cases, the results (summarized in Table 1) indicate that the adaptive lenses can correct the aberration almost completely over the correction area. However, only a small portion in the nearby of the corrected area respects the Marechal criterion for a well corrected system (RMS error of the aberration < $\lambda/14$ waves RMS). Nevertheless, the whole image benefits from the wavefront correction in both experiments. The average aberration error on the whole image was 0.33 waves RMS (central point) and 0.3 waves RMS (top-right point), while the portion of image being well corrected is 3.7% and 8.6% respectively.

Therefore, we conclude that the adaptive lens, in combination with the SPAM, allows one to correct the aberration in a chosen region of the sample, but residual aberration remains (in this case mainly defocus and astigmatism), away from the corrected location.

## 2.2 Aberrations measurement and correction in a light-sheet fluorescence microscope

Being able to measure and correct aberrations in a wide field configuration, the proposed device fits particularly well in the detection path of a Light Sheet Fluorescence Microscope (LSFM) [20]. This microscopy technique is based on the excitation of a single plane (light sheet) of a three-dimensional fluorescent sample. Widefield detection optics are used to collect light orthogonally to the excitation plane. The selective illumination of a single plane provides intrinsic optical sectioning capability, offering a powerful tool for the three-dimensional imaging of biological samples. LSFM is particularly well suited for imaging large, chemically cleared specimens, such as entire small animal's organs and tissues. When imaging chemically cleared samples, the mm to cm-sized specimens are placed in a high refractive index medium. This refractive index is approximately n≈1.4, for samples cleared in gel based media, such as Clarity [2], and n≈1.5 for samples cleared in solvents such as Benzyl Benzoate (BABB) and dibenzyl-ether (DBE) [21]. In many LSFM configuration the sample is placed in a glass cuvette, since the solvents would damage standard liquid immersion objective-lenses.

To demonstrate that the presented technique is compatible with the use of standard long working distance microscope objective lenses, we used a LSFM microscope based on illumination with a cylindrical lens (see Materials and Methods). The aberration correction module was placed in the detection arm of the microscope, right after the objective lens. Nonetheless, the light sheet thickness was relatively large (the beam waist was c.a. 7 μm) so to have a uniform light sheet thickness over the entire field of view, with negligible illumination aberrations. We initially tested the module on fluorescence nano-beads (Estapor F1-XC 010). These polystyrene spheres have an average diameter of 160nm, largely below the detection diffraction limit, and two maxima fluorescence peaks at 525nm and 560nm. A beads solution in water was embedded in a 1.5% phytagel in water solution and then poured in a Fluorinated Ethylene Propylene (FEP) tube, which has a refractive index close to that of water. The tube was then inserted in an Optical Glass square cuvette filled with distilled water (thickness of 2cm) for imaging. The system showed a strong spherical aberration induced by the interfaces between the air, the glass and the water contained in the cuvette. The aberration was stronger for off-axis fields, with a coma component. A summary of the wavefront measurements is shown in Table 2. The wavefront correction applied only in the central point led to a very small residual error (0.01waves) while it left a non-negligible coma component in the image corners (0.23waved RMS). The beads profile, full width at half maximum (FWHM) was reduced from 1.79 μm to 1.35 μm showing a considerable improvement in resolution. The beads profile in the corner of the image decreases to 1.57 μm. This result confirms what observed in the widefield measurement of the microscope slide: the correction is effective in one point (at the center) but leaves significant aberrations at the margin of the image.

**Table 2: Wavefront data relative to the fluorescent nano-beads acquisition.**

| | No AL FWHM (μm) | AL Corrected FWHM (μm) | No AL RMS error (waves) | AL Corrected RMS error (waves) |
|---|---|---|---|---|
| **Theoretical** | 1.13 | | | |
| **Center** | 1.79 | 1.35 | 0.54 | 0.01 |
| **Bottom left corner** | 2.30 | 1.57 | 0.69 | 0.23 |

We repeated the experiment on a biological sample, a chemically cleared adult zebrafish brain immersed in DBE. This sample shows strong autofluorescence, primarily in the blood vessels. The results are shown in Fig 4. The figure panels show the magnified details of two sample regions corresponding to the brain Optic Tectum (center) and to the Diencephalon (bottom-left).

For a fair comparison of the quality of the acquired images, we initially removed any adaptive lenses from the microscope (Fig. 4a). Then, we inserted the adaptive lenses for wavefront correction. As shown in Fig. 4a the correction in the central field leads to a clear contrast improvement over the overall image. The wavefront measurements (see Table 3) show that the aberration is almost completely corrected in the central field (0.05 waves RMS) while residual coma is present in the corners (0.27 waves RMS). Even if the fluorescence background is decreased in the entire field of view, improving the contrast of the acquired image, the resolution is not enhanced at the corners of the sample. An example is presented in Fig. 3f-g (showing a magnified detail of the red square in Fig. 3), where the correction with a single adaptive optics element is not sufficient to improve the resolution at the margins of the sample, and some brain vessels appear even more blurred that in the uncorrected case. A further strategy is required, to extend the correction to the entire field of view.

**Table 3: Wavefront error data relative the zebrafish brain sample (Fig. 3), corrected in the central region.**

| region | RMS (waves) |
|---|---|
| center | 0.05 |
| bottom left | 0.27 |
| mean in the full image | 0.22 |

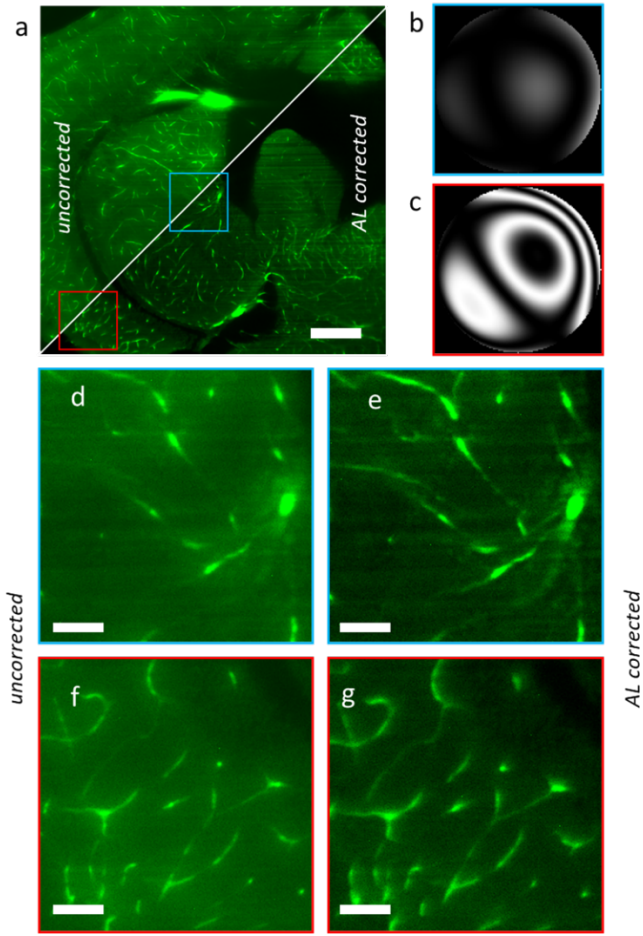

*Fig. 3. Light sheet microscopy acquisition of a chemically cleared adult zebrafish brain section, without and with AL correction. (a) Full image uncorrected (top-left triangle) vs corrected with the adaptive lens (bottom-right). (b) Wavefront aberration in the central region (blue in panel a). (c) Wavefront aberration at the left-bottom concern (red). (d-g) Magnified regions uncorrected (d, f) and corrected with the adaptive lens (e, g). Scalebar is 200 µm in a and 40 µm in b.*

**2.3 Correction on an extended area**

Altogether, the previous LSFM measurement suggest that the aberration contribution primarily comes from the refractive index mismatch, given by the cuvette. Its correction with a single adaptive optics element, leaves a residual field curvature and coma oriented towards the image center. This constitutes a general problem in the field of adaptive optics: the corrected area is limited to a small region, located around the correction point. Its size, called isoplanatic patch, depends on the volumetric aberration [16], [22], [23]. In astronomy, this limitation has been overcome by correcting the wavefront with a series of deformable mirrors in the so called Multi Conjugate Adaptive Optics (MCAO) [24] configuration. However, its implementation in microscopy has been so far inconvenient, due to the difficulties in using more deformable mirrors together with the lack of a wavefront measurement system [25].

Moreover, we verified the possibility of using multi conjugated adaptive optics for the present case. From our simulations in the case of a residual azimuthally oriented coma, such as the one presented in Fig. 4, even two-phase modulators with an infinite resolution would not be sufficient. In the case of using three phase modulators the maximum achievable correction area would be only 75% of the image. For the principle of phase conjugation, to achieve a perfect correction, the opposite volumetric structure of the aberration in the object space must be reported to the image space. This is not possible for a flat interface separating two regions with different refractive index with a finite number of phase modulators as for the presented case.

We therefore decided to put in place a multi-region anisoplanatic deconvolution [18]. We started from the measurement of the wavefront phase, which is provided by SPAM in multiple subregions over the entire image. In Fig. 4 the phases are shown in different areas (5x5 regions are presented), along with the measured RMS wavefront error. For each region, we calculated the 2D Fourier Transform of the wavefront and took its squared value, to generate the corresponding point spread function (PSF). Figure 4d shows that the PSF has a typical comet-like shape with an elongation that radially increases around the correction region. The acquired image was then divided in tiles, deconvolved with the corresponding PSF (with a custom Richardson-Lucy iterative approach), and recomposed to form a single, higher resolution, image [18].

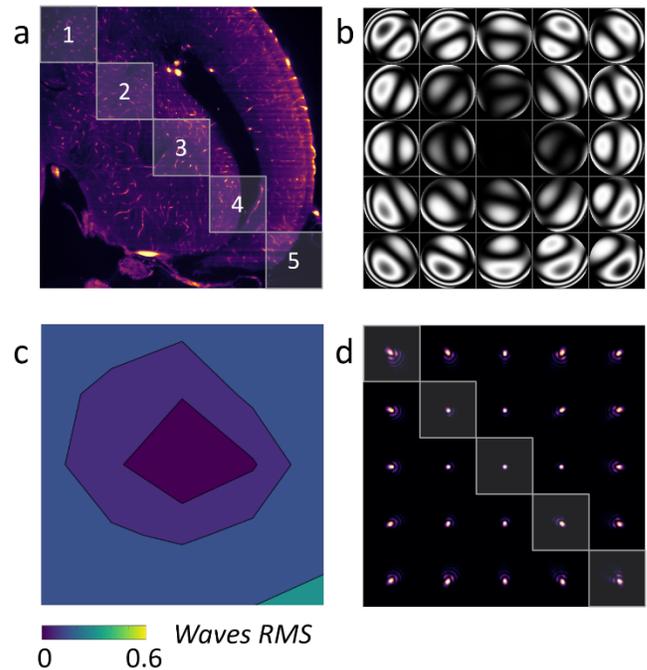

*Fig. 4. (a) LSFM acquisition of a zebrafish brain with AL correction in the image centre. (b) Wavefront measured in 5X5 subregions of the image. (c) RMS wavefront error (d) PSF generated for each subregion.*

The deconvolution successfully reconstructs the image in the entire field of view (Fig. 5), improving the image contrast and the resolution and facilitating the identification of single micro-vessels. Taking into consideration the image center (Fig. 5c), which is already corrected by the adaptive lens, we observe that the deconvolution does not significantly vary the image quality. The strength of the deconvolution is instead observable in the outer regions of the field of view (Fig. 5b and Fig. 5d), where the PSF has a strongly asymmetric and elongated shape. Each PSF, in fact, represents how a single point-source would appear in that given region: deconvolving with a wrong PSF may lead to the production of artifact. As an intuitive example, let us imagine to deconvolve the signal emitted by a bead located at the extreme corner of the image (as in Fig. 4d) with the -wrong- symmetric kernel measured in the central part. This operation would simply enhance each "feature" of the bead-comet but it will not remove its peculiar structure from the image, producing artifacts.

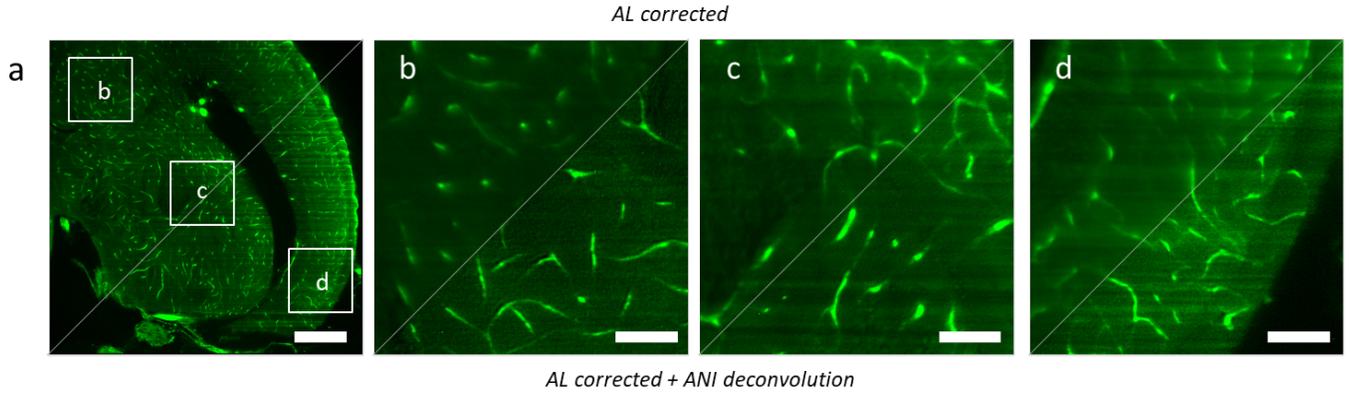

*AL corrected* (top) / *AL corrected + ANI deconvolution* (bottom)

*Fig. 5. Comparison between hardware correction with adaptive lens (AL) and the combination of hardware correction with anisoplanatic deconvolution (ANI). AL corrected is shown in the top left triangles, AL corrected with ANI deconvolution is shown in the bottom right triangle. (a) Full image of a zebrafish brain plane (b) Magnified region in the top left corner. (c) Magnified region in the center. (d) Magnified region in the bottom right corner. Scalebar is 200 µm in a and 50 µm in b-d.*

We can note the presence of horizontal lines in the deconvolved image, particularly in Fig. 5d. These are given by the known shadowing artifacts typical of light sheet microscopy, which have not been removed in this setup: light sheet is randomly attenuated along the propagation (light propagates from left to right in Fig. 5) by the absorbing and scattering sample, generating the lines. More importantly the visibility of the vessels and microvasculature is improved. We estimate the improvement in the image quality by calculating the contrast, here defined as the standard deviation of the intensity distribution of the image with respect to its mean. The original image has a contrast of 0.8, which, after isoplanatic deconvolution increases to 0.83, while for the anisoplanatic deconvolution, it increases to 0.93. The image is also affected by a significant distortion that, at the object corners, rises to 1.0 µm (see Supplementary text). Furthermore, the anisoplanatic deconvolution effectively compensates for this aberration too (Supplementary Fig.1). It is worth noting that a standard isoplanatic deconvolution would not successfully correct the deformations introduced in the image (Supplementaty Fig. 2): the knowledge of the region-dependent aberrations and the anisoplanatic deconvolution play a central role in high contrast and high-resolution reconstruction.

## 3. MATERIALS AND METHODS

### 3.1 Adaptive optics

We used a Multi Actuator adaptive lens, similar to the one described in [10]. These lenses rely on the use of ultrathin glass membranes bended by piezoelectric actuators. The lens is composed by two glass membranes with the inner space filled with a transparent liquid. It mounts two piezoelectric actuator rings, one on each face. Each ring is then segmented in 9 parts for a total of 18 singularly addressable actuators, so that aberrations up to the 4th order Zernike polynomials can be generated. The optical performances of the lenses used in these experiments are reported in the Supplementary Text.

### 3.2 Wavefront measurement and closed loop control

The wavefront measurement system is composed by a spinning wheel that, rotating, scans the pupil with an iris of 2.5 mm. The scan happens on 4 radial distances, with the angular positions of the iris being designed to avoid overlaps. The total wavefront points are 18. The system automatically adjusts the exposure time proportionally to the ratio R of the area of the sub-apertures and of the whole pupil (R =16). For the zebrafish samples, the exposure time with the fully open pupil was 40 ms and 640 ms for the sub-apertures. Hence, the total time for wavefront measurement was about 12 seconds. The spinning wheel is equipped with a large aperture used for image acquisition (Fig. 1) which only slightly limits the detection numerical aperture. The wavefront measurement was also used for the characterization of the deformable lenses necessary to operate the closed loop control. The characterization was performed poking each actuator individually and storing their wavefront deformation in a matrix (the so-called influence functions matrix). This operation was performed before the experiment using a phantom target in 9 positions of the field of view arranged in 3 x 3 equally spaced points. Then, in the correction process we could select the positions used for the wavefront correction.

### 3.3 Light-sheet fluorescence microscope

Light sheet fluorescence microscopy was performed on a custom system, similar that described in Ref. [27]. The system was based on a diode-pumped solid-state laser emitting at 473 nm, coupled in a single-mode optical fiber that delivered the light to the sample. The laser beam was collimated to a 2 mm waist and propagated through a cylindrical lens (f = 50 mm). The illumination power was 1.2 mW. A 10x, long working distance objective (Mitutoyo, NA = 0.28, WD = 34 mm) was used to collect the fluorescence signal. The objective, in combination with a tube lens and a long pass filter (FELH500, Thorlabs), formed the image on a CMOS camera (ORCA Flash 4.0, Hamamatsu), with a field of view of ca. 11mm×11mm. The sample was held in a glass cuvette of size 20mm x 20mm and it was moved with a 3-axis stage.

### 3.4 Anisoplanatic deconvolution

The SPAM permits the measurements of the PSF changes within different regions of the image. To exploit this accurate information, we decided to formulate an anisoplanatic deconvolution problem [18]. To begin, each acquired image is divided into overlapping squared regions (tiles). For our purpose, we chose the tile dimension to be $W \times W$, where $W = 410 px$. Each tile crops a portion of the image and is translated by steps of 82 pixels along both dimensions. By doing so, we obtain a total of $N \times N$ tiles, with $N = 21$. The number of tiles should be equal to the number of PSFs that we are able to estimate. Thus, we interpolate the 5x5 PSF map to match the number of total tiles of the image, 21x21. Then, each tile is deconvolved by its corresponding PSF with a custom Richardon-Lucy (RL) algorithm written in python. In our case, we take advantage of the convolution functions from the PyTorch library, running 50 RL-iterations. Once done, we recompose the final image by placing each deconvolved tile in its appropriate location, removing an external canvas of 32px to avoid the import of boundary artifacts. As a reassembling rule, the overlapping regions of the tiles where averaged by counting the number of tile contributions in each single pixel. This produces the reconstruction of the anisoplanatically deconvolved image.

### 3.5 Zebrafish brain

The adult zebrafish brain was prepared as follows (in accordance with the protocol n. 513/2018-PR, authorized on 4/7/2018 by the Italian Health Ministry). The fish sacrificed by over-anaesthesia in a tricaine solution of 500 mg/l. When opercular movements stopped and all reflex reactivity was lost, the fish was rapidly placed in a cut made in a sponge within a Petri dish containing chilled tricaine solution. The brain was quickly and carefully removed and placed overnight in 4% paraformaldehyde on a shaker (at 4°C). The brain was subsequently processed as described in Ref. [26].

To reveal the origin of the moderate/high autofluorescence recorded at brain level we performed fixation by perfusion. The fish was placed in a tricaine solution (150 mg/l). When opercular movements stopped and all reflex reactivity was lost, the fish was placed in a sponge (the Petri dish contained 150 mg/l chilled tricaine solution). The heart was exposed, and a smoothed 30 Gauge needle inserted into the ventricle. A saline solution was flushed (1-2 ml, at 1 ml/min) and when fish gills whitened, a fixative solution (10 ml) was used to perfuse. The brain was then carefully removed and placed overnight in 2% paraformaldehyde on a shaker (at 4°C). The brain was subsequently processed as described above. Light sheet fluorescence microscopy reveled absence of autofluorescence in the brain (not shown), indicating that autofluorescence derived from blood-containing brain vessels.

## 4. CONCLUSIONS

We have demonstrated that the use of deformable lenses in combination with anisoplanatic deconvolution increases the size of the corrected area on aberrated microscopy images. This allowed us to combine the deformable lens with the aberration detection system, based on a spinning sub-pupil, that enables wavefront measurement directly from the detected image. The SPAM system provides information about the aberrations in every point of the field of view. This is particularly important because it gives quantitative information about the aberration structure generated by the sample. Furthermore, in the presented cases, we evaluated the size of the isoplanatic patch after wavefront correction. It is worth noting that, with respect to all the literature relative to wavefront sensorless aberration correction in microscopy, we were able to measure the size of the well corrected patch in the case of the aberration correction with a single wavefront modulator. The results show that, even if the part of the image with diffraction limited features is extremely small (from 4% to 8%, Table 1), on average, the whole image benefits from the correction (from 0.78 waves RMS down to about 0.3 waves RMS, Table 1). The fact that the wavefront phase is measured over the entire field of view has another major impact. The residual aberration, out of the isoplanatic patch, can be quantified and corrected by deconvolution, providing high resolution reconstruction over the entire field of view.

Finally, the measurement and correction of aberrations using the integrated SPAM and deformable lens module, can be of easy installation on many types of microscopes.

## Acknowledgments

CNR-IFN was supported by Office of Naval Research global (ONRG) and by Air Force Research Laboratory (AFRL) with grant GRANT12789919. Politecnico di Milano has received funding from LASERLAB-EUROPE (grant agreement no. 871124, European Union's Horizon 2020 research and innovation programme) and from H2020 Marie Skłodowska-Curie Actions (HI-PHRET project, 799230).

We are grateful to Dr Giulia Petrillo for helping in the zebrafish brain preparation.

## Disclosures

S. Bonora acknowledges a financial interest in Dynamic Optics srl.